# Verifying Non-friendly Formal Verification Designs: Can We Start Earlier?

Bryan Olmos, Rheinland-Pfalzische Technische Universität Kaiserslautern-Landau, Infineon Technologies AG, München, Germany (*bryan.olmos@infineon.com*)

Daniel Gerl, Infineon Technologies AG, München, Germany (*daniel.gerl@infineon.com*)

Aman Kumar, Infineon Technologies AG, München, Germany (*aman.kumar@infineon.com*)

Djones Lettnin, Infineon Technologies AG, München, Germany (djones.lettnin@*infineon.com*)

*Abstract—* **The design of Systems on Chips (SoCs) is becoming more and more complex due to technological advancements. Missed bugs can cause drastic failures in safety-critical environments leading to the endangerment of lives. To overcome these drastic failures, formal property verification (FPV) has been applied in the industry. However, there exist multiple hardware designs where the results of FPV are not conclusive even for long runtimes of model-checking tools. For this reason, the use of High-level Equivalence Checking (HLEC) tools has been proposed in the last few years. However, the procedure for how to use it inside an industrial toolchain has not been defined. For this reason, we proposed an automated methodology based on metamodeling techniques which consist of two main steps. First, an untimed algorithmic description written in C++ is verified in an early stage using generated assertions; the advantage of this step is that the assertions at the software level run in seconds and we can start our analysis with conclusive results about our algorithm before starting to write the RTL (Register Transfer Level) design. Second, this algorithmic description is verified against its sequential design using HLEC and the respective metamodel parameters. The results show that the presented methodology can find bugs early related to the algorithmic description and prepare the setup for the HLEC verification. This helps to reduce the verification efforts to set up the tool and write the properties manually which is always error-prone. The proposed framework can help teams working on datapaths to verify and make decisions in an early stage of the verification flow.**

## I.  INTRODUCTION

Digital designs, especially those involving complex datapath operations like arithmetic functions, often begin with a prototype developed in a high-level language such as C, C++ or SystemC [1]. These prototypes serve as a starting point for exploring various implementations of algorithms during the early stages of design. Engineers use these prototypes to validate the specification as well as to explore the architecture before to implement the algorithm at the RTL, making refinements to optimize power, performance, and area.

The former studies have highlighted the applicability of Formal Verification (FV) in critical applications such as automotive, but it may not scale well to larger designs due to the state explosion problem, especially in non-friendly formal verification designs such as multipliers, FPU (Floating Point Units), filters and designs implementing complex algorithms [2]. Furthermore, FPV does not scale well for arithmetic units. Most of the approaches give up when the bit width increases. One alternative is High-Level Equivalence Checking (HLEC) which offers a technique to verify the functional equivalence between different design representations. It is more intuitive for verification engineers as HLEC focuses on comparing two different representations of the same design instead of proving properties.

However, errors during the concept phase are usually not detected until the verification of the RTL design. Additionally, each stage of the design process can introduce bugs when a manual approach is used. Other approaches such as simulation cannot provide enough reliability for safety-critical designs. Furthermore, the verification time consumes more than 50% time of the total verification process [3]. For these reasons, this paper introduces a methodology called MetaHLEC which expands the use of FV for exhaustively verifying datapaths, like arithmetic functions, by reusing C/C++ reference implementations through HLEC. The methodology uses an automation framework to formalize specifications for use in safety-critical, requirement-driven development flows, reducing manual efforts that could cause structural errors. Formalizing specifications using metamodeling techniques can minimize misinterpretation, enabling automation of parts of the design and verification flow, saving time and improving the quality of verification results and design. For example, 10 assertions related to the design of an FPU were proven at the software level within 4.9s and the HLEC yields full proof in 40.9s. This is a big advantage compared with the consumed time to model equivalent SVA properties that incorporate floating point multiplication as well as expected state space explosion, where no comparable model checking results could be obtained. Additionally, other efforts of constrained random simulation cannot prove the absence of bugs for all input combinations.

The main contributions of this paper include:

• We present MetaHLEC which is a new framework for the automation of HLEC and the metamodeling of data path designs for the generation of properties for the verification of untimed algorithmic descriptions written in C++, which is used in combination with commercial formal verification tools.

• Analysis and verification of 6 types of datapath designs such as Unsigned Single-Instruction Multiple Data (SIMD) Multiplier, Floating Point Multiplier, Quadratic Fractional Polynomial, Pipelined Unsigned Division, Finite Impulse Response (FIR) Filter, and Error Correcting Codes.

The rest of this paper is organized as follows. In Section II, we introduce the challenges of the designs under verification (DUVs) and the related work. The proposed methodology and metamodel are described in Section III. Section IV presents the verification results including a comparison with FPV. At last, we conclude this work in Section V.

II. BACKGROUND

*A. Verification Challenges in Datapath Designs*

The verification of datapath designs presents significant challenges due to their inherent complexity and the intricate mathematical operations they involve. These operations include parallel data processing, precision requirements, pipelining, and error detection and correction. The process of verifying these circuits and algorithms must account for various input values, which can be challenging as the number of possible combinations related to the data width. It can quickly become unmanageable for formal verification. Taking in account the feasibility, scalability and the resource utilization, a design under verification can be seen as friendly or unfriendly to formal methods [2][4]. On the one hand, formal friendly datapath designs prioritize high concurrency and low sequential depth, along with control logic elements and low complexity data transformations. These designs often involve parallel data processing and are characterized by simplified bus protocols and straightforward data processing. On the other hand, non-friendly datapath involve designs implementing complex bus protocols, such as AXI (Advanced eXtensible Interface), OCP 2.1 (Open Core Protocol), PCI (Peripheral Component Interconnect); designs with complex arithmetic units and designs with a a large state space. In this work, we focused on the challenges related to following designs:

- Unsigned Single-Instruction Multiple Data (SIMD) Multiplier: Complex data paths and control logic make time-consuming ensuring the correct behavior across multiple data elements and handle corner cases and boundary conditions.
- Floating Point Multiplier: The IEEE 754 standard establishes precision requirements, rounding modes, and exception handling to ensure the correct implementation of floating-point multiplication, including handling of denormalized numbers such as NaNs and infinities[5]. It requires rigorous formal reasoning and verification which makes it complex writing properties for the verification.
- Pipelined Unsigned Division: Verifying the correct operation of the pipelined division algorithm across multiple stages and ensuring precise handling of quotient and remainder calculation is a challenging task due to the complex pipelining, data hazards, and control logic associated with the division process.

- Quadratic Fractional Polynomial: Formal verification of quadratic fractional polynomials can be challenging due to the intricate mathematical operations involved, including multiplication, addition, and division of fractional coefficients.
- Finite Impulse Response (FIR) Filter: Formal verification of FIR filters is challenging due to the intricate signal processing algorithms, including convolution, coefficient multiplication, and delay elements.
- Error Correcting Codes (ECC): It is challenging due to the complex encoding and decoding algorithms, as well as the need to guarantee the correctness of error detection and correction. Its verification includes properties relates to parity generation, syndrome calculation, and error correction capability.

*B. Related Work*

Ludwig et al. is [6], [7] use a simulation-based approach for verifying SystemC models. However, formal verification of SystemC remains very challenging because of overhead by object-oriented structures and simulation-specific semantics [8], [9]. Additionally, the approach of property-driven design also does not ease the verification problem of datapaths. Thus, [11] motivates the usage of C/C++ over SystemC for designing algorithmic circuits. Another approach for utilizing software verification tools to verify RTL implementations is proposed as translating RTL into cycle-accurate C. Mukherjee et al. [12] translate Verilog designs into cycle-accurate ANSI-C programs using their tool V2C [13]. Their analysis shows significant verification speedups especially in datapath-intensive designs. However, this approach cannot be used in an early stage of the design. Additionally, their tool V2C is limited to Verilog [14]. Qurat-ul-Ain et al. [15] made improvements to correct translation errors from V2C. They, however, use the verified C program to perform High-Level Synthesis (HLS) and obtain an optimized RTL design. However, the HLS design is verified with simulation. The approaches described in [14], [16][17][18] allow formal proofs between C/C++ references and RTL implementations. However, all of them rely on golden C/C++ reference models without providing means to verify their correctness according to specification.

III. METHODOLOGY AND APPLICATION

The proposed methodology in Figure 1 can be applied once a design specification is finalized. In this work, the specification is associated into the metamodel of Section III.A . Usually, the design process starts with a design or concept engineer implementing a C or C++ function that represents the functionality or partial functionality of the design. As the C and C++ languages do not define any notion of time, the algorithm does not contain information about the latency or throughput of the implementation. However, it can be used efficiently to provide conclusive results about our algorithm taking in account that the properties in C can be verified on a few seconds as shown in Section IV. This first step is called Algorithm Verification and is a precondition to continue the second step called Implementation Verification. Algorithm Verification takes the C/C++ algorithm reference and assertions generated from the verification plan to perform BMC — for example the open-source tool CBMC [9]. Successful model checking proves the correctness of the algorithm exhaustively, thereby qualifying the algorithmic implementation as a golden reference model. Afterwards, the algorithm is implemented in RTL using the scheduling and throughput constraints from the specification. In Implementation Verification, the C/C++ reference algorithm verified in the previous step and the RTL implementation are the main inputs to a commercial HLEC tool. It proves the equivalence of the DUV to the algorithm with respect to scheduling information from the specification.

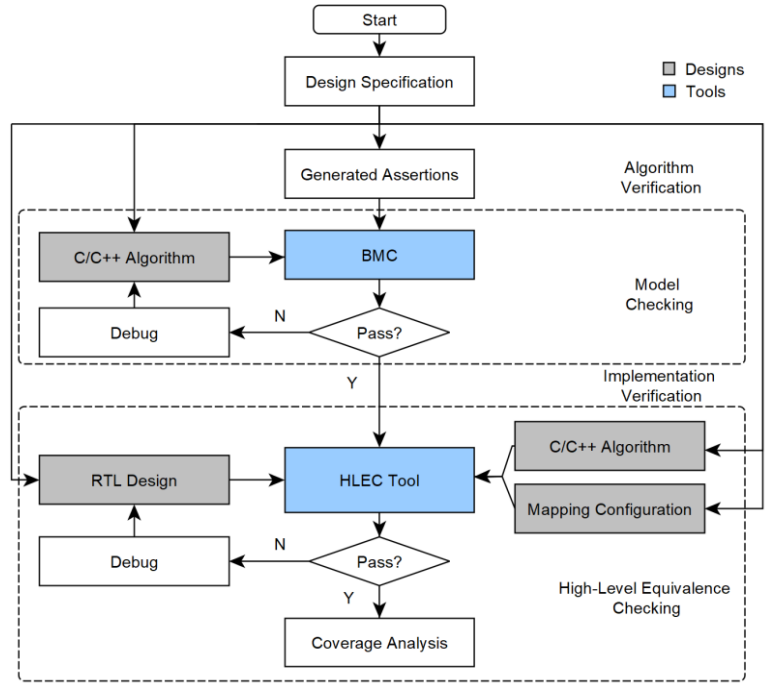

Figure 1. Overview of the verification methodology

*A. Metamodel MetaHLEC*

As described in the previous section, the verification process is conducted in two verification cycles which will be referred to as Algorithm Verification and Implementation Verification. The needed information for these verification cycles is added into the metamodel of Figure 2. The root node MetaHLEC contains the name of the DUV as its only attribute. It can have one or multiple Requirement nodes, which represent functional requirements that the C/C++ algorithm has to fulfill, as well as a Mapping node that contains all information needed to connect the C/C++ algorithm to the RTL implementation. This Metamodel consists of two main nodes:

- Requirement node: It contains the information for the verification of the algorithmic description and consists of a description, a name, and an ID for unique identification inside our internal specification management system. As part of the formalization of our algorithm, it is expressed as a cause-and-effect behavior, such that an action shall follow if a guarding expression occurs. For example, a guard can define action to take if a division by 0 occurs in the algorithm implementing the Pipelined Unsigned Division. Expressions can be seen as a structure consisting of literal, operators, and variable; for example, *input_value != 0 ( input_value: literal, operator: !=, variable:0)*. An operator can be associated with other operators in case of more complex properties needs to be defined.

- Mapping node: It contains the additional information for the HLEC and specifies the names of the implementation module (Imp_name) and algorithm function (Spec_name). It contains a Clock and Reset signal, which are only defined at RTL. Input and Output are the signals that have to be mapped. For a complete model, at least one input and one output need to be declared. In the inputs and output signals, it needs to be defined the name, the delay, the port size, and the sign interpretation for the port mapping. The mapping will be only active if the expression of the Condition is fulfilled. It is necessary for pipelined designs with variable delays, then the HLEC is verified only in the clock cycles that fulfill the condition. Additionally, our metamodel also considers a pipeline stalling condition, which can be achieved by output composition with *Stalling*. Furthermore, environmental conditions are implemented via Constraint, which similarly to guards and actions can be mapped to a management system via an ID, description, and name. Helper specifies the assertions to guide the RTL verification process.

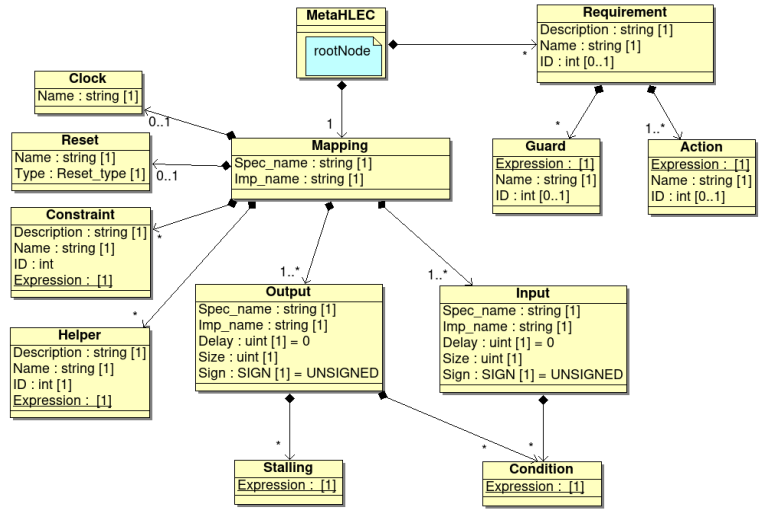

Figure 2. Reduced diagram of the metamodel

While the model can be represented as UML-diagram, its instantiation will be stored in structured XML. The instantiation of the model is guided by a Graphical User Interface (GUI).

### B. Automation MetaHLEC

Based on the XML file associated to the introduced metamodel, the automation framework was developed. It minimizes the user interaction and structural errors. The algorithm specification will be formalized by transforming into guarding and action expressions that define the combinational behavior. The behavioral description is transformed into assertions in the target-language C to perform functional verification of the algorithmic design representation. Information about the design implementation consisting of clocks, resets and ports as well as operating and stalling conditions are used to generate the verification setup script. The fully automated translation from model to target code allows quick verification adaption to changes in functional and timing requirements while minimizing human error during environment setup.

Figure 3 shows the overall generation process where the metamodel information and MAKO templates [23] are used to generate the verification elements such as a C harness containing the properties, the runscripts and the SV wrapper used to compare the C code and the RTL design. Figure 4 shows an example of the SVA properties used for FPV (Formal Property Verification) and its equivalent generated property for C verification considering a scalable division algorithm for unsigned integers, which adds one stage to the pipeline for every data bit. A pipeline for a 16-bit division for example, consists of 16 stages and delays calculation of the quotient by 17 clock cycles. Additionally, the design considers an undefined division by 0 via a flag. If an invalid operation occurs the quotient shall be set to all ones.

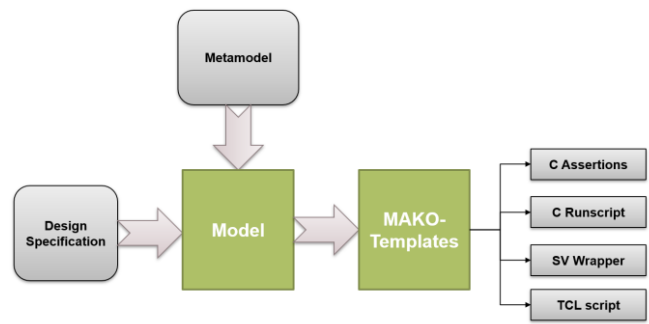

Figure 3. The design specification is captured in a model instance based on the defined parameters of the metamodel

```
                SVA properties for division                                    Generated C assertions for division
1  // Divisor is 0                                          1  //Requirement 0: Division by zero
2  property div_by_0;                                       2    if( ( b_i == 0 ) ){
3  ( (b_in == '0 ) |->                                      3      __CPROVER_assert(( divide_by_0_o == 1 ), "Flag set");
4  ##(num_stages +1) divide_by_0_out && quotient_out == '1  4
5  );                                                       5      __CPROVER_assert(( ( ~ quotient_o ) == 0 ), "Division by zero");
6  endproperty: div_by_0                                    6    }
7                                                           7
8  // Divisor not 0                                         8
9  property div_quotient_out;                               9    //Requirement 1: Divisor not 0
10 ( ( b_in != '0 ) |->                                     10   if( ( b_i != 0 ) ){
11 ##(num_stages +1) !divide_by_0_out && quotient_out ==    11     __CPROVER_assert(( divide_by_0_o == 0 ), "Flag clear");
12 ($past(a_in, (num_stages +1))                            12
13 / $past(b_in, (num_stages +1))));                        13     __CPROVER_assert(( quotient_o == ( a_i  / b_i ) ), "Quotient");
14 endproperty: div_quotient_out                            14   }
```

Figure 4. Comparison of properties for the verification of the Pipelined Unsigned Division.

## IV. RESULTS

Table I shows the main results for three evaluated open-source designs. For the evaluation, the timeout was set up to 24h. Additionally, due to time-consuming modelling of equivalent SVA-properties that incorporate floating point multiplication as well as expected state space explosion, no comparable model checking results could be obtained. The results show that HLEC checking can verify designs which cannot be handle with a Formal Property Verification (FPV) approach.

Table I. Verification time results for open-source designs

| Design Under Test | Proposed Methodology | | | Normal Approach |
|---|---|---|---|---|
| | *CBMC* | *EDA HLEC Tool* | *Total* | *EDA FPV Tool* |
| Unsigned Single-Instruction Multiple Data (SIMD) Multiplier (16 bits) [19] | 1.1 s | 194.4 s | 195.5 s | Timeout [a] |
| Floating Point Multiplication [20] | 4.9 s | 40.9 s | 45.8 s | Unknown |
| Quadratic Fractional Polynomial (Fractional Width = 7) [21] | 0.03 s | 138.3 s | 138.33s | Timeout |

[a.] *4-bit and 8-bit multiplication: 505.3 s.*

Afterwards, a scalable division algorithm for unsigned integers was verified. HLEC is proving the design for data widths up to 52 bit within the specified timeout of 24 h. While model checking obtained faster results for data widths lower than 8 bit as visible in Figure 5, the runtime obtained with HLEC rises at a much slower rate than with property checking indicating better scalability of the proposed methodology.

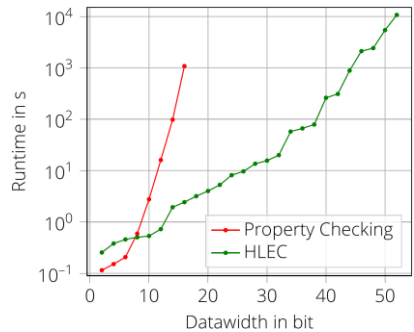

Figure 5. Proof runtime for scalable data width of the division operators

The results of the verification of a discrete filter FIR that is used in data processing [16] are shown in Figure 6. For the 8-bit implementation in the range of orders 1 to 31 HLEC showed an average runtime decrease by a factor of 177. The 16-bit implementation could only be proven up to an order of 21 with the property checking setup without timeout after 24 h.

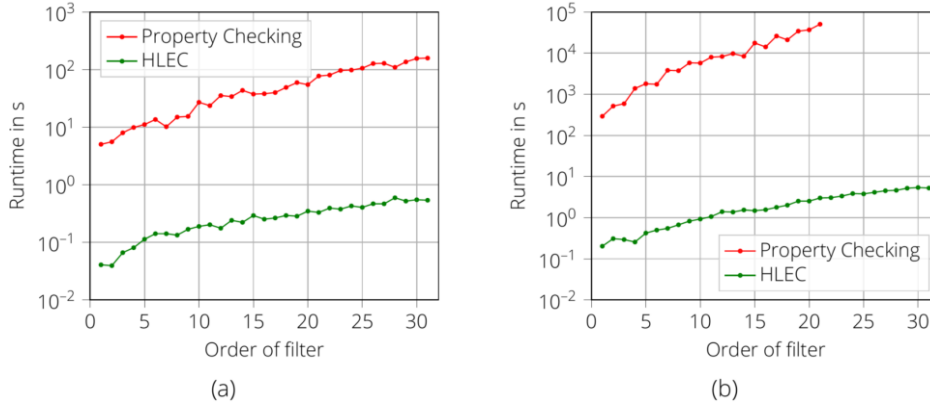
Figure 6. Comparison of HLEC and property checking runtimes for filter with (a) 8-bit and (b) 16-bit precision for data and coefficients

Finally, the implementation of an industrial design of ECC (Error Correcting Codes) was evaluated. As depicted in Table II an ECC with 32-bit data-input with single-error correction and double-error detection was proven at a third of the runtime that the linearity approach described achieved with only 5.7 s. For the 64-bit data width with 3/4 correction/detection, CBMC still performed acceptably at 1 second, but the EDA HLEC tool experienced a timeout, and the Formal Property Verification (FPV) tool took significantly longer at 16016.93 seconds. Other implementations with larger datawidths and larger error correction capabilities, however, were not verifiable with the developed methodology. The abstraction level chosen for this IP was too high in order to obtain conclusive results.

Table II. Verification time results for open-source designs

| Datawidth | Correction/Detection | Runtime CBMC | Runtime EDA HLEC Tool | Runtime FPV |
|---|---|---|---|---|
| 32 | 1/2 | 0.97 s | 5.7 s | 17.2 s |
| 64 | 3/4 | 1 s | Timeout | 16016.93s |

All results shown in this section were obtained EDA tool vendors for C2RTL application for HLEC and FPV application for property checking, CBMC-5.73.0 [17] and Z3-4.12.0 [18].

## V. CONCLUSION

The automation framework generates a script that sets up the verification environment, maps the ports taking mapping conditions and delays into account, executes the proof, and samples coverage information. Thereby, the model-based approach allows for optimized input mapping minimizing the additional complexity delay operations add. The developed methodology was applied to several IPs. The runtime results show the methodology's potential within data processing applications over FPV approaches. Its exhaustive proofs make it preferable over dynamic verification approaches as these can often never achieve full coverage of all states in the design. In combination with the automation framework that allows integration into requirement-driven flows, the methodology is especially suited for safety-critical applications. The verification time for Unsigned Single-Instruction Multiple Data (SIMD) Multiplier with 4 operation modes, a 32 bits floating point multiplication unit with 4 operations modes and Quadratic Fractional Polynomial were 195.5 s, 45.8s and 138.3s, respectively. Additionally, a scalable division algorithm for unsigned integers was verified. HLEC is capable to verify designs for data widths up to 52 bits within the specified timeout of 24 h. While model checking obtained faster results for data widths lower than 8 bits, the runtime obtained with HLEC shows a better scalability of the proposed methodology. With respect to the verification of a discrete filter FIR that is used in data processing, for the 8-bit implementation in the range of orders 1 to 31 HLEC showed an average runtime decrease by a factor of 177. The 16-bit implementation could only be proven up to an order of 21 with the property checking setup without timeout after 24 h, while the proposed methodology could provide exhaustive proof up to order 31. Additionally, 32-bit implementation could be proven within 54s up to an order of 64 while FPV remained inconclusive for any order.

Finally, the implementation of an industrial design of ECC (Error Correcting Codes) was evaluated. However, ECCs implementations with larger data widths and larger error correction capabilities were not verifiable with the developed methodology. The reason is that the abstraction level chosen for this IP was too high in order to obtain conclusive results.

VI. ACKNOWLEDGMENT

This work has been developed in the project VE-VIDES (project label 16ME0243K) which is partly funded within the Research Programme ICT 2020 by the German Federal Ministry of Education and Research (BMBF).